\PassOptionsToClass{12pt}{revtex4-1}
\documentclass[manusript]{aastex631}
\usepackage[T1]{fontenc}
\usepackage[]{graphicx,amsmath,bm}

\shortauthors{Clarkson \& Kontar}

\begin{document}

\title{Magnetic Field Geometry and Anisotropic Scattering Effects on Solar Radio Burst Observations}

\author[0000-0003-1967-5078]{Daniel L. Clarkson}
\affiliation{School of Physics \& Astronomy, University of Glasgow, Glasgow, G12 8QQ, UK}

\author[0000-0002-8078-0902]{Eduard P. Kontar}
\affiliation{School of Physics \& Astronomy, University of Glasgow, Glasgow, G12 8QQ, UK}

\begin{abstract}
The fine structures of solar radio bursts reveal complex dynamics in the corona, 
yet the observed characteristics of these sub-second bursts are additionally complicated by radio wave scattering in the turbulent solar corona. We examine the impact of anisotropic turbulence in radio-wave propagation simulations with non-radial magnetic field structures in shaping the morphology, time-characteristics, and source position of fine structures. The apparent sources are found to move along the direction of the magnetic-field lines and not along the density gradient, whereas the major axis of the scattered source is perpendicular to the local magnetic field (the scattering anisotropy axis). Using a dipolar magnetic field structure of an active region, we reproduce observed radio fine structure source motion parallel to the solar limb associated with a coronal loop and provide a natural explanation for puzzling observations of solar radio burst position motions with LOFAR. Furthermore, the anisotropy aligned with a dipolar magnetic field causes the apparent source images to bifurcate into two distinct components, with characteristic sizes smaller than in unmagnetized media. The temporal broadening induced by scattering reduces the observed frequency drift rate of fine structures, depending on the contribution of scattering to the time profile. The findings underscore the role of magnetic field geometry and anisotropic scattering for the interpretation of solar radio bursts and highlight that anisotropic scattering produces more than a single source.
\end{abstract}


\section{Introduction}

Solar radio bursts are frequently observed from the solar corona and
are often associated with the acceleration of non-thermal electrons in solar flares 
\citep{1972ARA&A..10..159W, 1985ARA&A..23..169D, 2008A&ARv..16....1P}. 
The radio sources evolve over time and through space as non-thermal electrons drive Langmuir waves 
that produce radio emission due to the plasma emission process, emitted close to the local plasma frequency or its harmonic. 
The radio waves experience substantial scattering effects as the emission propagates through an anisotropic turbulent medium. 
In such an environment, the dominant contribution to scattering arises due to plasma density 
inhomogeneities close to the inner scale \citep{2023ApJ...956..112K}. The process incites broadening of the observed time profile \citep[e.g.][]{2020ApJ...905...43C} and apparent size \cite[][]{2017NatCo...8.1515K,2021A&A...645A..11M}, combined with an often radial displacement of the apparent source position \citep[e.g.][]{2018ApJ...868...79C,Kontar_2019,2022ApJ...925..140G}. Consequently, the scattering contribution has a major implication for radio-burst imaging---the location and size of the emitting region is not observed.

For Type III sources, the motion of the electron beam with velocity $v_\mathrm{b}$ through space causes the emission 
to evolve in both time and frequency simultaneously as $\mathrm{d}f/\mathrm{d}t \propto v_\mathrm{b}$, 
manifesting as a drift in dynamic spectra \citep[e.g.][]{1973SoPh...29..197A}. 
A frequency drift over time is also evident for fine structures at a smaller drift rate. For example, \cite{2018SoPh..293..115S} measure Type IIIb striae drift rates around 30 kHz s$^{-1}$ at 30 MHz embedded within a Type III burst with a bulk drift rate of 10 MHz s$^{-1}$. Given that electron beam speeds associated with the drift rate of the burst envelope are typically a fraction of light speed $\sim c/3$ \citep[e.g.][]{1959AuJPh..12..369W, 1985srph.book..289S}, then the striae drift rate that is orders of magnitude lower must require a different driving mechanism. As a result, fine structure drift rates have been linked to density fluctuations and propagating magnetohydrodynamic waves \citep{2018SoPh..293..115S, 2018ApJ...861...33K}, and recently associated with the drift of Langmuir waves in space, which is dependent on the coronal temperature \citep{2021NatAs...5..796R}. However, the scattering-induced broadening of the time profile in dynamic spectra may dilute the observed burst drift rate as suggested by \cite{Clarkson_2021}, which is particularly significant for narrow bandwidth fine structures such as spikes and Type IIIb striae. Since the observed drift rate serves as an indication of the source motion through space, then an important consequence of the scattering component is to influence the interpretation of the driver. In the case of fine structures, their observed scattering-affected drift rates would estimate decreased Langmuir wave group velocities and coronal temperatures, unless the scattering component is decoupled.

Recent investigation into the influence of anisotropic radio-wave scattering on apparent source evolution have considered radially symmetric magnetic fields \citep{Kontar_2019, 2020ApJ...905...43C, 2020ApJ...898...94K}. In such geometry, the anisotropy of the density fluctuation spectrum causes a scattering rate that is stronger perpendicular to the radial direction. 
The sources are broadened asymmetrically, and the photons preferentially escape radially away from the Sun parallel to the field, manifesting as the radial displacement of the source at sub-second scales. 
However, recent fine structure observations of Type IIIb striae and spikes 
using the LOw Frequency ARray (LOFAR) between 30--45 MHz revealed non-radial, 
fixed frequency source motion over time, 
attributed to anisotropic scattering in the environment 
with a non-radial magnetic field (a coronal loop-like structure) \citep{2023ApJ...946...33C}. Presently, such magnetic field configurations 
have not been combined with anisotropic scattering simulations 
in order to explain such features.

In this study, we investigate the role of anisotropic scattering on radio source motion in non-radial magnetic fields, and the resulting fine structure morphology in dynamic spectra. We extend the scattering model by \cite{Kontar_2019} to account for the magnetic field configuration of coronal loops approximated by a dipole, 
and investigate the directivity of the escaping radiation through analysis of the simulated image centroid motion, and the drift rates in dynamic spectra via convolution with a scattering function. In section 2, we describe the simulation approach and assumptions. In section 3, we present the numerical simulation results, and in section 4, we investigate the scattering effect on fine structure drift rates in simulated dynamic spectra.

\begin{figure}[b!]
\centering
    \includegraphics[width=0.5\textwidth]{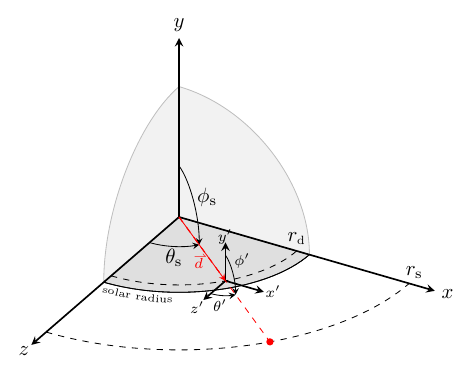}
    \caption{Sun centered coordinate system $\bm{r}=(r,\theta,\phi)$ for an observer located along the positive $z$-axis where $\theta=0\arcdeg$ and $\phi=0\arcdeg$. The radio source is shown at position $(r_\mathrm{s}, \theta_\mathrm{s}, \phi_\mathrm{s})$ The off-centered dipole with axes $(x^\prime, y^\prime, z^\prime)$ is shown at position $\bm{d}=(r_\mathrm{d}, \theta_\mathrm{d}, \phi_\mathrm{d})$ with the magnetic field components described by $(r^\prime, \theta^\prime, \phi^\prime)$.}
    \label{fig:coords}
\end{figure}

\section{Ray-tracing Simulation Set-up}\label{sec:sim_setup}

\begin{figure*}[htb!]
    \epsscale{1.}
    \plottwo{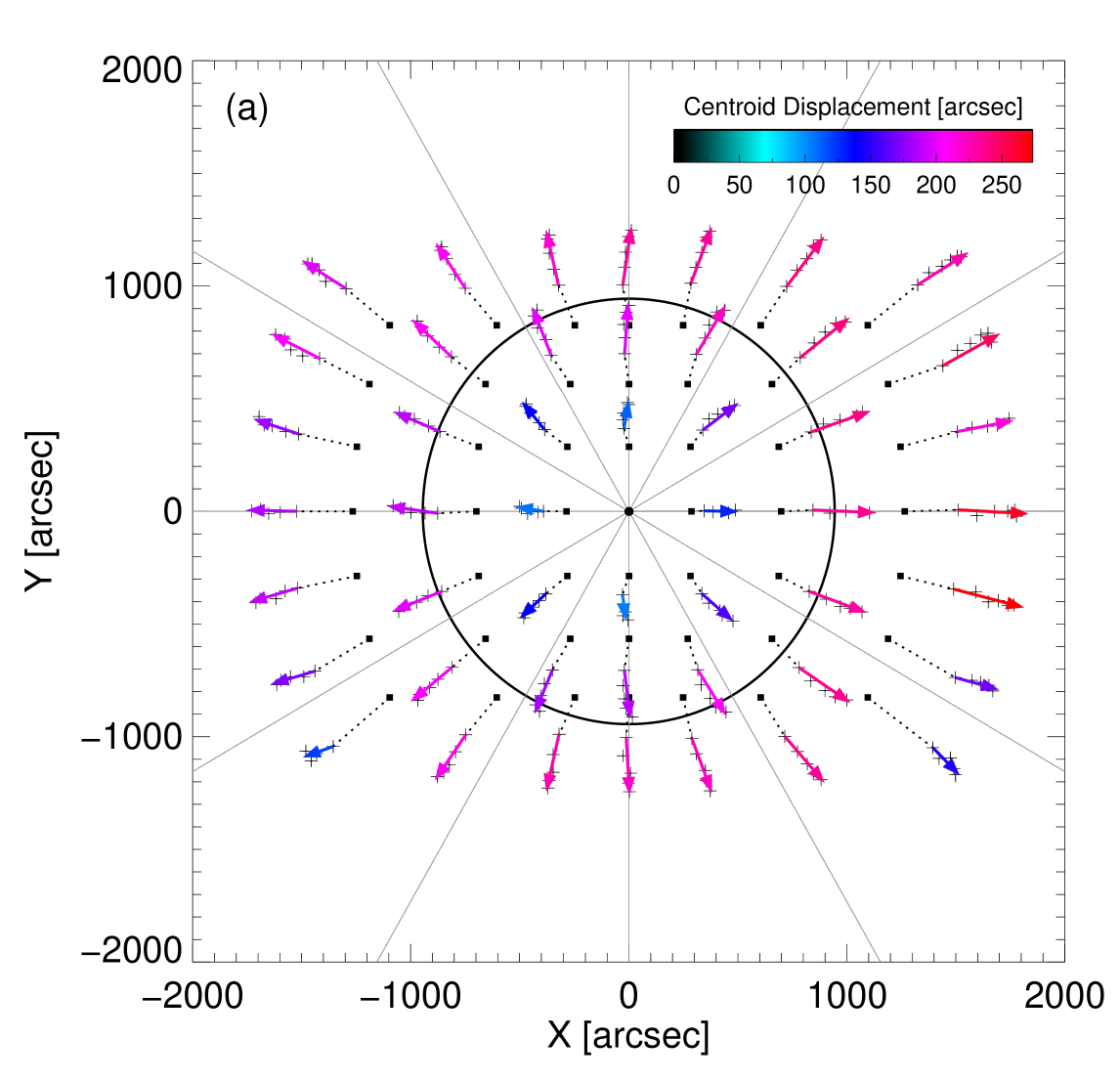}{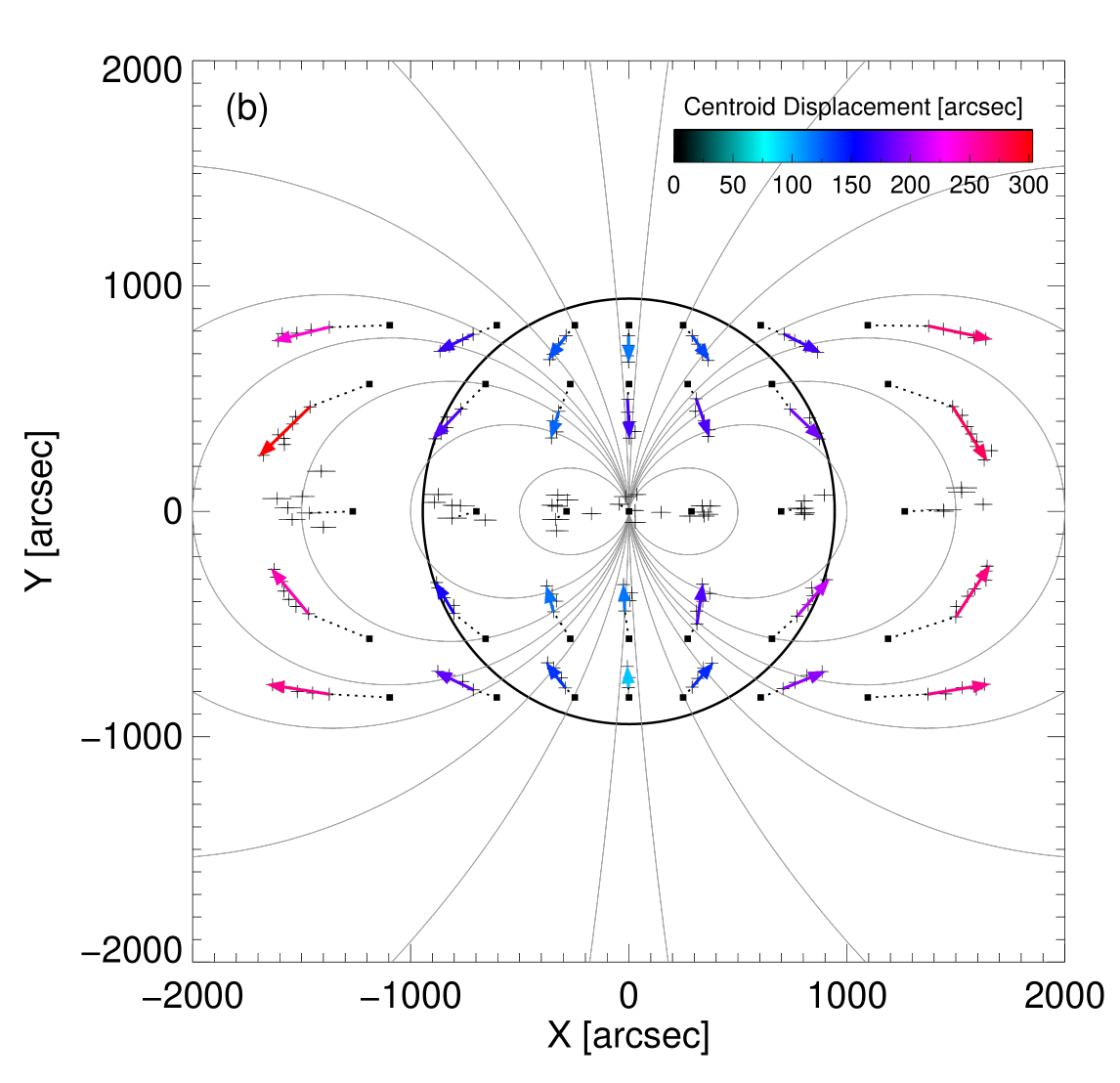}
    \caption{Centroid trajectories across the FWHM for sources distributed across the solar disk in a (a) radial and (b) dipole magnetic field, with $\alpha=0.2$. The dipole is located at the solar center. The vectors are colored according to the centroid displacement distance. The black squares give the source injection locations, and the dotted line connects the corresponding centroid locations (black plus symbols).}
    \label{fig:fullSun_radial_dipole_vectors}
\end{figure*}

The numerical code is based on that described in \cite{Kontar_2019} for the observations by \citet{Clarkson_2021, 2023ApJ...946...33C}. We consider initial source heights $r_\mathrm{s}$ at a distance of $r_\mathrm{s}=1.80$~R$_\odot$ and $r_\mathrm{s}=1.75$~R$_\odot$, corresponding to a plasma frequency of $f_\mathrm{pe}=28.7$~MHz and $f_\mathrm{pe}=32$~MHz. Assuming an emission ratio of $f=1.1f_\mathrm{pe}$ (fundamental emission, see polarization arguments by \citet{Clarkson_2021}) which is conserved throughout the simulation, the emission frequencies are $f=31.6$~MHz and $f=35.2$~MHz.
The photons are injected as an initial delta function with isotropic wavevector distribution. The turbulence profile is described by the spectrum-weighted mean wavenumber of the density fluctuations and given in \cite{2023ApJ...956..112K} (see their equation 14), where $\alpha$ quantifies the anisotropy of the density fluctuation spectrum as $\alpha=q_\parallel/q_\perp$, and $q$ is the wavenumber of the density fluctuations. When $\alpha=1$, the density fluctuation spectrum is isotropic, and anisotropic when $\alpha<1$ with axial symmetry with respect to the guiding magnetic field. We introduce a multiplicative scaling factor $\eta$ to increase or decrease the turbulence level (note the factor 1/4, 1/2, 2, 4 used 
by \citet{2023ApJ...956..112K} to explain the variety of solar radio observations).

Figure \ref{fig:coords} shows the Sun-centered coordinate system with axes $(x, y, z)$ and spherical coordinates $(r,\theta,\phi)$ where $\theta$ is the azimuthal angle from the \textit{z}-axis in the range 0 to $2\pi$, and $\phi$ is the polar angle between $0$ and $\pi$. We approximate the presence of a coronal loop with a dipole magnetic field model, 
where the dipole center is shifted to a location given by $\bm{d}=(r_\mathrm{d},\theta_\mathrm{d},\phi_\mathrm{d})$, with the magnetic field components defined with respect to axes $(x^\prime, y^\prime, z^\prime)$ with transformations $x^\prime=x-x_\mathrm{d}$, $y^\prime=y-y_\mathrm{d}$, and $z^\prime=z-z_\mathrm{d}$. We assume that the radio source is a point source given by the Sun-centered coordinates $(r_\mathrm{s},\theta_\mathrm{s},\phi_\mathrm{s})$ where the observer lies along the $+z$-axis such that the $x,y$ axes form the plane-of-sky. The off-centered dipole is represented via $(r^\prime, \theta^\prime, \phi^\prime)$ where
\begin{equation}
    \begin{split}
        r^\prime &= \sqrt{r^2 + r_\mathrm{d}^2 - 2rr_\mathrm{d}\left(\sin{\theta}\sin{\theta_\mathrm{d}}\cos{\left(\phi-\phi_\mathrm{d}\right)} + \cos{\theta}\cos{\theta_\mathrm{d}}\right)} \, , \\
        \theta^\prime &= \arccos{\left(y^\prime/r^\prime\right)} \, , \\
        \phi^\prime &= \arctan{\left(x^\prime, z^\prime\right)} \, .
    \end{split}
\end{equation}
The magnetic field vector is given by
\begin{equation}\label{eq:dipole_eq}
    \bm{B}(r^\prime, \phi^\prime)=\frac{B_0}{(r^\prime)^3}\left(2\cos{(\phi^\prime)}\mathbf{\hat{r}^\prime} + \sin{(\phi^\prime)}\boldsymbol{\hat{\phi}^\prime}\right),
\end{equation}
where $B_0$ is a normalization constant. Despite being a simple model for the field above an active region, the approximation is used in the corona far from the photosphere \citep[e.g.][]{1984SoPh...94..249K, 1999SoPh..185..157R}.

\begin{figure*}[htb!]
    \epsscale{1.2}
    \plotone{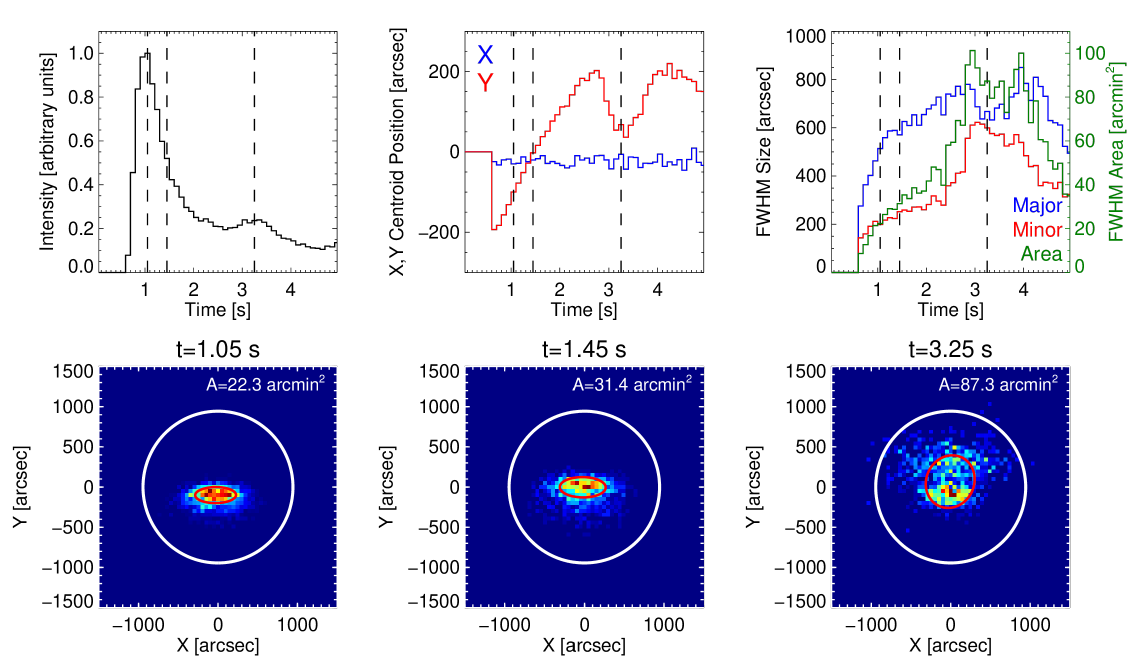}
    \caption{Simulation results for an injected source of $N=10^6$ photons at $f=35.2$~MHz for $\theta_\mathrm{s}=0\arcdeg$ and $\phi_\mathrm{s}=-5\arcdeg$, turbulence profile scaling factor $\eta=0.5$ and anisotropy $\alpha=0.2$. The dipole located at $\theta_\mathrm{d}=0\arcdeg$, $\phi_\mathrm{d}=0\arcdeg$, and $r_\mathrm{d}=0.9$ R$_\odot$. The top panels show the intensity time profile, the $x$ (blue) and $y$ (red) centroid positions, the FWHM major (blue) and minor (red) sizes, and the FHWM area (green) respectively. The dashed lines mark the times of the images (weighted 2D histograms) presented in the lower panels from left to right, generated from the observed photon $x,y$ locations with a binsize of 50 arcsec. The centroid position, sizes, and area are determined from a 2D Gaussian fit to the images. The FWHM area is shown by the red oval.}
    \label{fig:sim_images_TH_FI_0}
\end{figure*}

\section{Numerical Simulation Results}

The photons experience scattering until a radius where the scattering rate becomes negligible. The photons are then collected and back-projected from the observer location to generate an $x,y$ photon position map \citep[see][for details]{Kontar_2019}. The observed photon locations are binned into a weighted 2D histogram with a binsize of 50 arcsec. The weighting $w$ of each photon $i$ is given as $w_i=e^{-\tau_a}$ where $\tau_a=\int{\gamma(\mathbf{r}(t))}\:\mathrm{d}t$, $\gamma=(\omega_{\mathrm{pe}}^2/\omega^2)\gamma_c$ is the free-free absorption rate, and $\gamma_c$ is as defined as in \cite{Kontar_2019}.
The 2D images are fit with an elliptical Gaussian of the form
\begin{equation}\label{eq:2d_gauss}
    I(x,y)=I_0\exp\left(-\frac{x^{\prime2}}{2\sigma_\mathrm{x}^2} - \frac{y^{\prime2}}{2\sigma_\mathrm{y}^2}\right),
\end{equation}
where $I_0$ is the maximum intensity, $\sigma_x,\sigma_y$ are the major and minor axis lengths, and
\begin{equation}
    \begin{aligned}
        x^\prime = (x-x_c)\cos{T} - (y-y_c)\sin{T},\\
        y^\prime = (x-x_c)\sin{T} + (y-y_c)\cos{T},
    \end{aligned}
\end{equation}
where $T$ is the ellipse rotation angle clockwise from the \textit{x}-axis. Equation \ref{eq:2d_gauss} provides the centroid locations $(x_c,y_c)$ of the observed emission used to indicate the source location, full-width at half-maximum (FWHM) major and minor sizes as $S_\mathrm{x,y}=\sigma_\mathrm{x,y}2\sqrt{2\ln{2}}$, and FWHM area as $A=\pi/4\cdot S_\mathrm{x}S_\mathrm{y}$.

\subsection{Radial and Dipole Magnetic Field}

\begin{figure*}
    \epsscale{1}
     \plotone{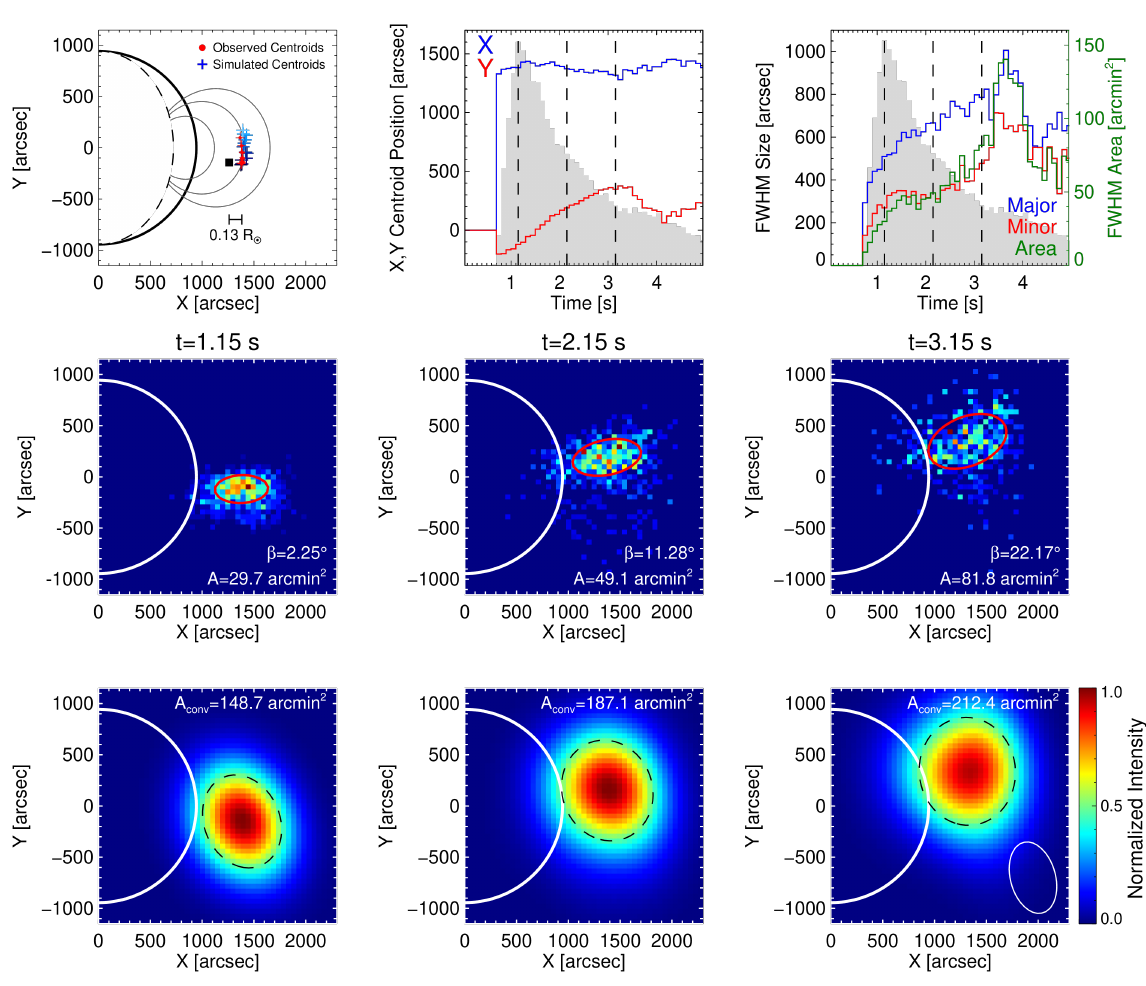}
    \caption{Simulation results for an injected source of $N=10^6$ photons at $f=35.2$~MHz for $\theta_\mathrm{s}=50\arcdeg$ and $\phi_\mathrm{s}=95\arcdeg$, with the dipole located at $\theta_\mathrm{d}=50\arcdeg$, $\phi_\mathrm{d}=90\arcdeg$ and $r_\mathrm{d}=0.7$~R$_\odot$, and turbulence profile scaling factor $\eta=0.5$ with anisotropy $\alpha=0.2$. In the top row, the left panel shows the simulated centroids (plus symbols) in the sky-plane with time across the FWHM increasing from dark to light blue. The observed centroids of a radio spike from \cite{Clarkson_2021} are shown by the red symbols, corrected for average ionospheric refraction as in \cite{2022ApJ...925..140G} (see equation 6). The black square gives the initial source location, with the distance indicator corresponding to the sky-plane distance between the photon injection point and the first observed centroid. The center and right panels show the centroid time profiles, the FWHM major and minor axis sizes, and FWHM area, overlaid onto the normalized intensity time profile (grey). The dashed lines mark the times of the images (weighted 2D histograms) presented in the center panels from left to right. The centroid position, sizes, and area are determined from a 2D Gaussian fit to the images. The FWHM area is shown by the red oval, and the angle between the fitted ellipse major axis and the $x$-axis is given by $\beta$. The lower panels show a convolution of each image with a 2D Gaussian to mimic a Low Band Antenna beam of LOFAR in the outer configuration with a $\sim3.5$~km baseline, shown in the lower right corner. The black dashed oval marks the 50\% intensity level with the FWHM convolved area given by $A_\mathrm{conv}$.}
    \label{fig:LOFAR_obs_comp}
\end{figure*}

To demonstrate the spatial evolution of radio sources in the presence of a non-radial magnetic field, Figure \ref{fig:fullSun_radial_dipole_vectors} shows the centroid trajectories measured from images across the FWHM time profile for simulations with initial sources distributed across the solar disk. Panel (a) uses a spherically symmetric radial magnetic field, and panel (b) uses a dipole field located at the solar center. In the radial case, the projected source motion into the sky-plane is radially outwards from the Sun, with the largest component of motion visible for sources closer to the limb. This is also apparent in the dipole case. However, there is additional variation of the source trajectories and displacement distance depending on the source polar angle $\phi_\mathrm{s}$, with the absence of any clear trajectory for sources located at $\phi_\mathrm{s}=0\arcdeg$.

Figure \ref{fig:sim_images_TH_FI_0} shows the results of a single simulation for a source located at $\theta_\mathrm{s}=0\arcdeg$ and $\phi_\mathrm{s}=95\arcdeg$ ($5\arcdeg$ below the equator), and a solar disk centered dipole at $r_\mathrm{d}=0.9$~R$_\odot$. Throughout the FWHM duration of the main peak, the source shifts along the field direction (which here is projected vertically in the sky-plane) by $\sim160$~arcsec over $0.65$~s. The bulk of this motion ($100$~arcsec) occurs throughout the decay phase. The FWHM linear size is larger in the \textit{x}-direction by a ratio of 2.2 and 2.5 at the peak and decay, respectively. The FWHM area increases almost at a constant rate from $14$ to $22$ and $32$~arcmin$^2$ at the start, peak, and end of the FWHM period. Near $3$~s, the major size stops increasing, with a sharp increase in the minor size due to the arrival of the echo \citep{2020ApJ...898...94K} as seen in the lightcurve, superimposed onto the main burst component. This causes the source location to jump to a lower position in the sky plane as the image is now composed of both the primary and reflected components, which consequently causes the measured source area to be enlarged, and of different shape and orientation. As the echo begins to shift in the same manner as the primary component, the source shifts along the $y$-direction, and the area begins to reduce as photons from the main burst component are no longer visible.

\subsection{Comparison with Observations of Spikes and Striae}

Figure \ref{fig:LOFAR_obs_comp} presents the simulation results for a configuration similar to Figure \ref{fig:sim_images_TH_FI_0}, yet with $\theta_\mathrm{s}=\theta_\mathrm{d}=50\arcdeg$ and $r_\mathrm{d}=0.7$~R$_\odot$. For the observed event, the active region is located at approximately $50\arcdeg$ from the observers line-of-sight (LOS) \citep{Clarkson_2021}. The simulated and observed source centroids are in agreement with the start location, and their trajectory across the sky-plane. From the simulations, we see a projected radial distance of $0.13$~R$_\odot$ between the photon injection location and the source location at the start of the rise phase. The simulated FWHM peak area is $A=29.7$~arcmin$^{2}$ compared to the observed spike peak area in \cite{Clarkson_2021} of $A_\mathrm{obs}=200$~arcmin$^{2}$ at $34.5$~MHz, with different orientation. Convolving the simulated images with a 2D Gaussian to mimic the point spread function (PSF) of a LOFAR Low Band Antenna beam using the outer configuration with a baseline of $\sim3.5$~km \citep{2013A&A...556A...2V} on the date of observation (15 July 2017 at 11:00 UT) increases the simulated peak area to $148.7$~arcmin$^2$ (scaling to $34.5$~MHz to match the observations gives $154.8$~arcmin$^2$, increasing as $1/f^2$).

\begin{figure}
    \centering
    \includegraphics[width=0.5\textwidth]{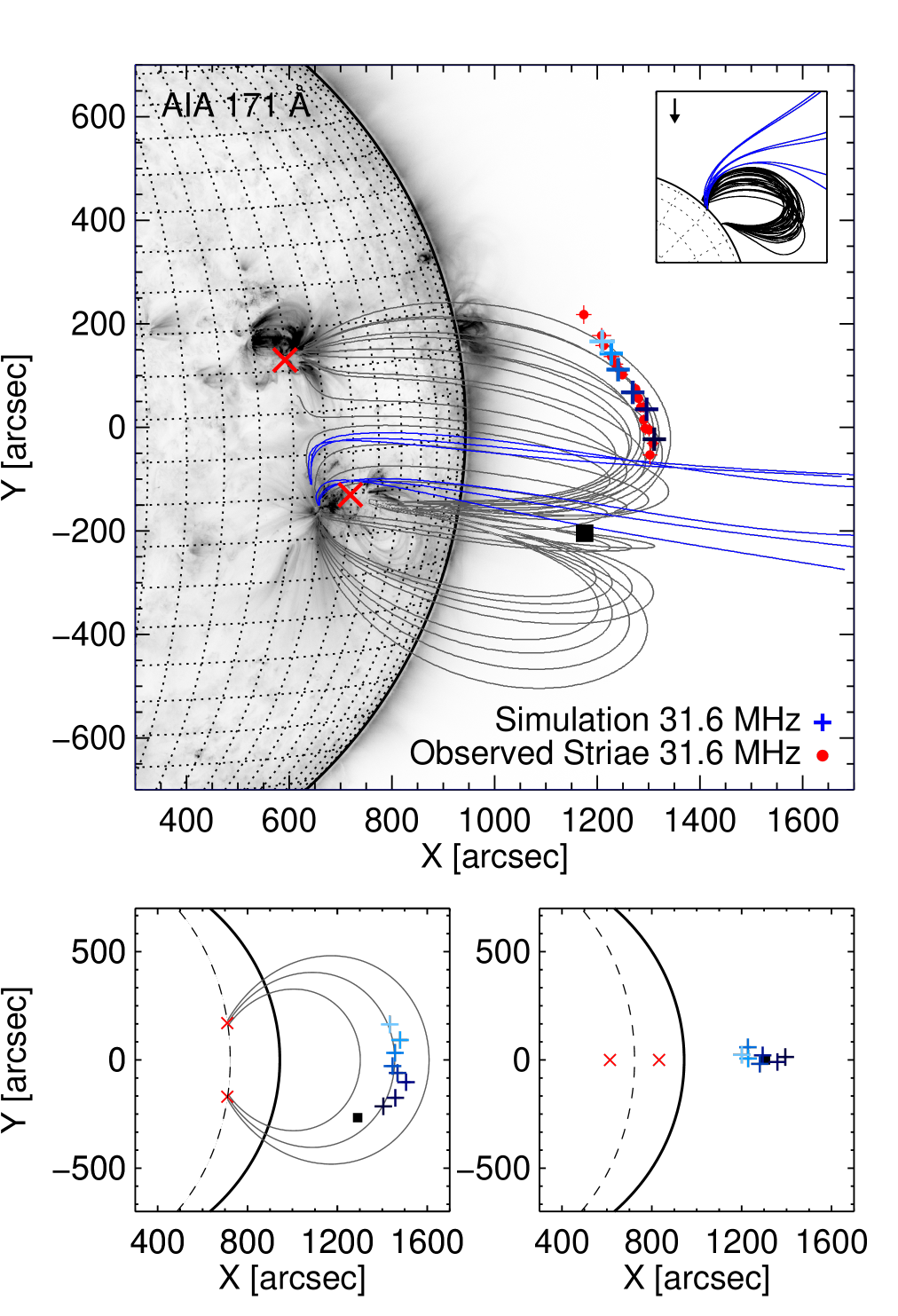}
    \caption{FWHM Centroid motion across the \textit{xy}-plane for an initial source injected at 31.6 MHz at $\theta_\mathrm{s}=50\arcdeg$ and $\phi_\mathrm{s}=100\arcdeg$ (black square, $10\arcdeg$ below the equator) with $\eta=0.5$ and $\alpha=0.2$. The top panel overlays an SDO/AIA $171$~\r{A} image from 15 July 2017 at 11:20 UT and a potential-field source-surface extrapolation of closed (grey) and open (blue) field lines with the simulated centroids (blue crosses, with time progressing from dark to light) where the dipole has been rotated anti-clockwise by $\psi=40\arcdeg$ to align the dipole footpoints (red crosses) with active regions in the AIA image. The inset shows the PFSS model viewed from the solar pole, with the arrow showing the view from Earth. The closed red circles show the observed Type IIIb striae centroids from an observation at $31.6$~MHz at the time of the AIA image \citep[][see Figure 3e]{Clarkson_2021} and are corrected for ionospheric refraction as in Figure \ref{fig:LOFAR_obs_comp}. The lower panels show a rotation of $\psi=0\arcdeg$ and $\psi=90\arcdeg$, respectively.}
    \label{fig:rdi09_Fisrc-5_gammaRot}
\end{figure}

In Figure \ref{fig:rdi09_Fisrc-5_gammaRot}, we show that the source trajectory can vary depending on the orientation of the dipole with respect to the solar surface. Acknowledging that a rotation of the source position around the Sun is equivalent to an opposite rotation of the observer, one can show the effective rotation of the dipole around the normal to the solar surface by rotating the photon position- and wave-vectors to the disk center by $-\theta_\mathrm{s}$, performing a rotation by an angle $\psi$ around the \textit{z}-axis, followed by a rotation to a desired source angle $\theta_\mathrm{s}$. The lower panels show the default configuration at $\theta_\mathrm{d}=50\arcdeg$, and an additional rotation anti-clockwise about the normal to the solar surface by $\psi=90\arcdeg$. The source trajectories vary in each case due to the anisotropic scattering-induced motion along the field direction, showing that radial motion towards the Sun is possible for certain configurations. The upper panel presents a rotation anti-clockwise about the normal to the solar surface by $\psi=40\arcdeg$ and placed such that the field lines that connect the sky-plane region of interest emerge from the solar surface aligned with the two active regions in the AIA image. The simulated source trajectories at 31.6 MHz now have a component of motion towards the solar limb, matching that of observed Type IIIb striae presented in \cite{Clarkson_2021} at the same frequency with a $0.24$~R$_\odot$ separation between the emission location and the first centroid of the rise phase.

\subsection{Apparent Source Bifurcation}

\begin{figure*}
    \epsscale{1.15}    
    \plotone{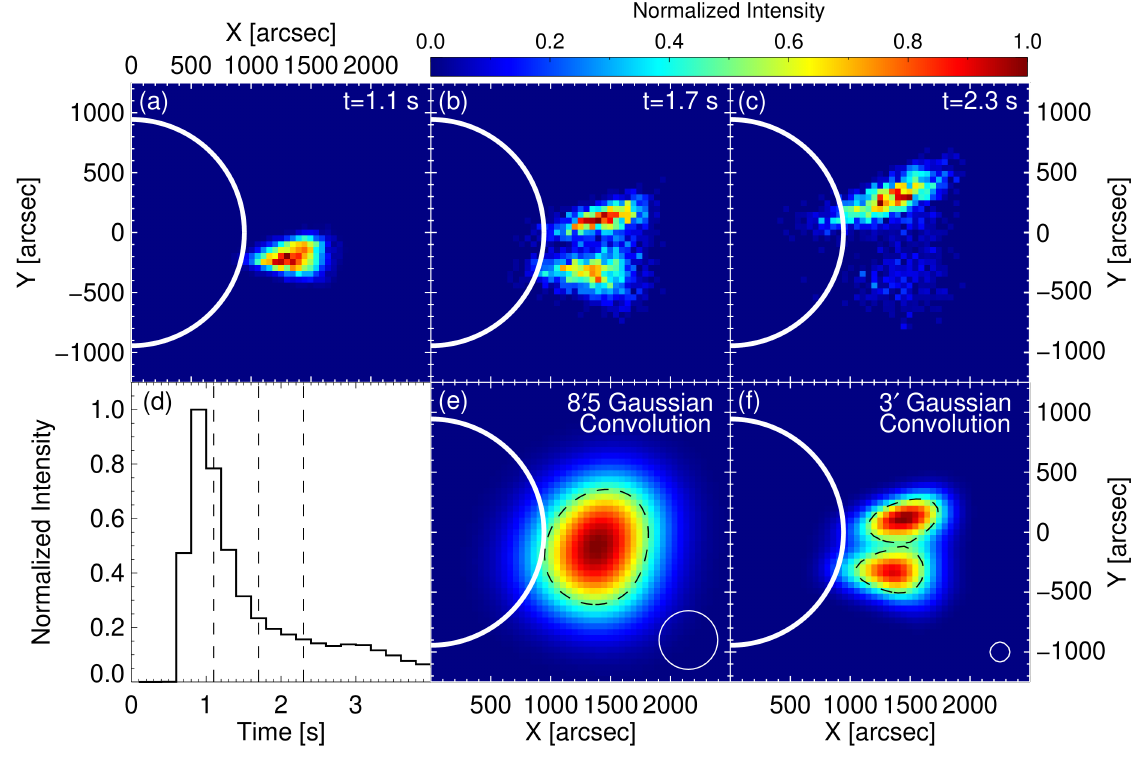}
    \caption{Simulated images (a-c) presented as weighted 2D histograms for a source and dipole located as in Figure \ref{fig:LOFAR_obs_comp} with a dipole depth of $r_\mathrm{d}=0.9$~R$_\odot$, using $10^6$ photons and anisotropy $\alpha=0.1$. Each image corresponds to the times shown in panel (d) by the vertical dashed lines. Panels (e,f) show the sources at $t=1.7$~s convolved with a 2D Gaussian of FWHM size $8.5$ matching that of LOFAR at 35 MHz and $3$~arcmin (allowing the two components to be resolved), represented by the white circular contours in the lower right corners. The black dashed lines show the $50\%$ contour levels.}
    \label{fig:split_source}
\end{figure*}

For a similar configuration used in Figure \ref{fig:LOFAR_obs_comp}, but with an anisotropy factor of $\alpha=0.1$ (Figure \ref{fig:split_source}), the imaged source appears as a single source in the vicinity of the emission site at the peak time, yet 0.8 s after the peak the apparent source branches into two distinct components. One component shifts upwards in the sky-plane along the field direction, while the other shifts downwards, with both components of similar size and a separation distance of $\sim0.4$ R$_\odot$. Towards the end of the decay phase $1.4$~s after the peak, the brighter source has shifted farther away, whilst the lower source is significantly fainter. The lower panels of Figure \ref{fig:split_source} show the simulated images at $t=1.7$~s convolved with a 2D Gaussian of different sizes, again to mimic an observing instrument point spread function. Panel (e) shows that a convolution with a Gaussian of size 8.5 arcmin is too large to resolve the two sources, whilst panel (f) shows a smaller Gaussian size of 3 arcmin enables the two sources to be spatially resolved at the $50\%$ intensity level, according to the Rayleigh criterion.

\subsection{Observed Characteristics with Source Polar Angle}

\begin{figure}
\centering
    \includegraphics[width=0.5\textwidth]{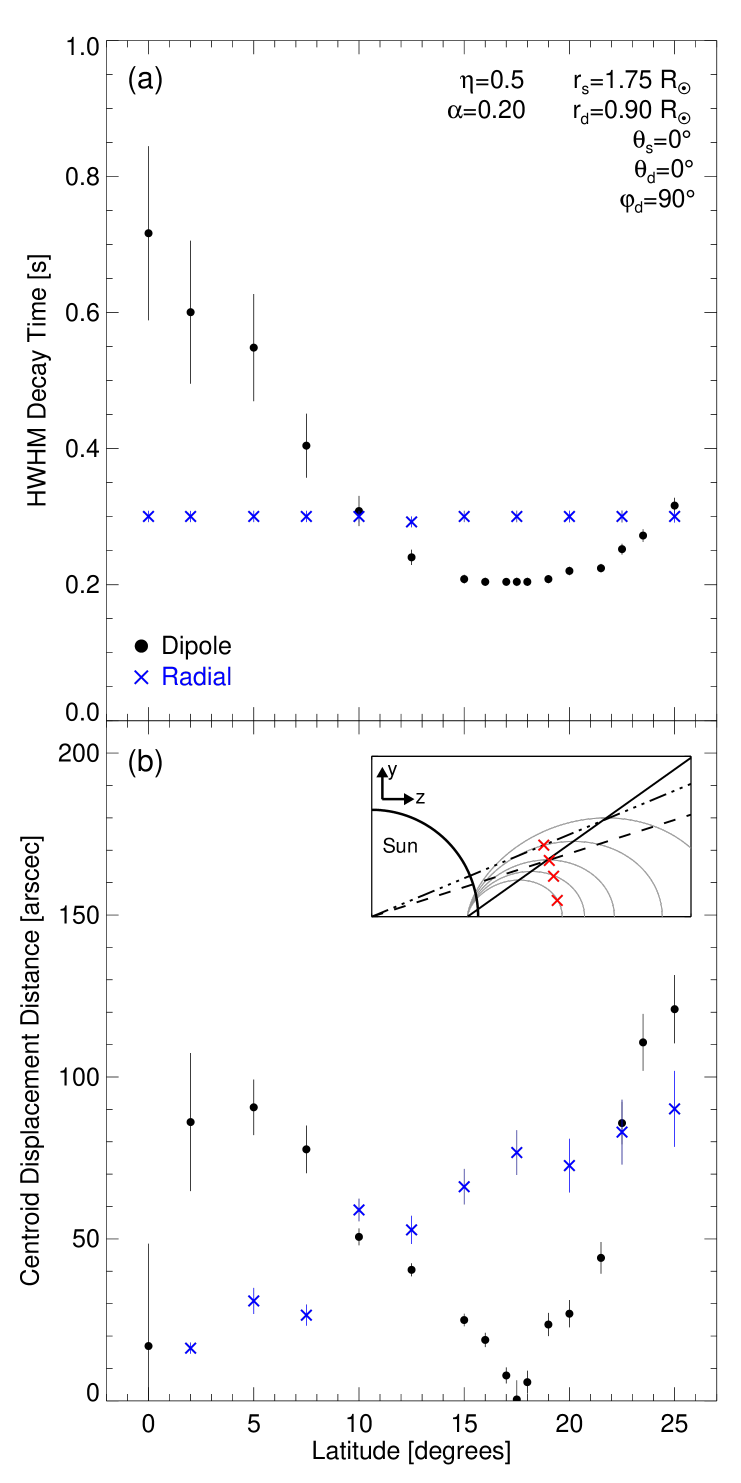}
    \caption{Simulated (a) decay times and (b) centroid displacement distance, measured from the peak to the FWHM level for sources at polar angles between $\phi_\mathrm{s}=65\arcdeg-90\arcdeg$ (up to $25\arcdeg$ from the equator) with $N=10^5$. In all cases, the dipole was offset to $r_\mathrm{d}=0.9$~R$_\odot$, at $\theta_\mathrm{d}=0\arcdeg$ and $\phi_\mathrm{d}=90\arcdeg$, with the sources azimuthal angle also $\theta_\mathrm{s}=0\arcdeg$. The turbulence profile is scaled by a factor of $\eta=0.5$ with anisotropy $\alpha=0.2$. The blue and black data represents the cases for radial and dipole magnetic fields, respectively. The decay uncertainty at lower $\phi$ results from lower photon counts (two orders of magnitude less than those near $17.5\arcdeg$). The inset of panel (b) displays the $zy$-plane and the source $\phi$ locations at 5, 12.5, 17.5, and 22.5 degrees (red crosses) overlaid with the dipole field lines. The solid line gives the angle of $35.3\arcdeg$ where each field line reaches a maximum height. The dashed line represents an angle of $17.65\arcdeg$ from the solar center where the displaced dipole reaches a maximum height for that source location, and the dash-dotted line marks an angle of $22.5\arcdeg$ from the solar center where the radial direction matches the field line direction.}
    \label{fig:decay_shift_fi_src}
\end{figure}

\begin{figure}
\centering
    \includegraphics[width=0.5\textwidth]{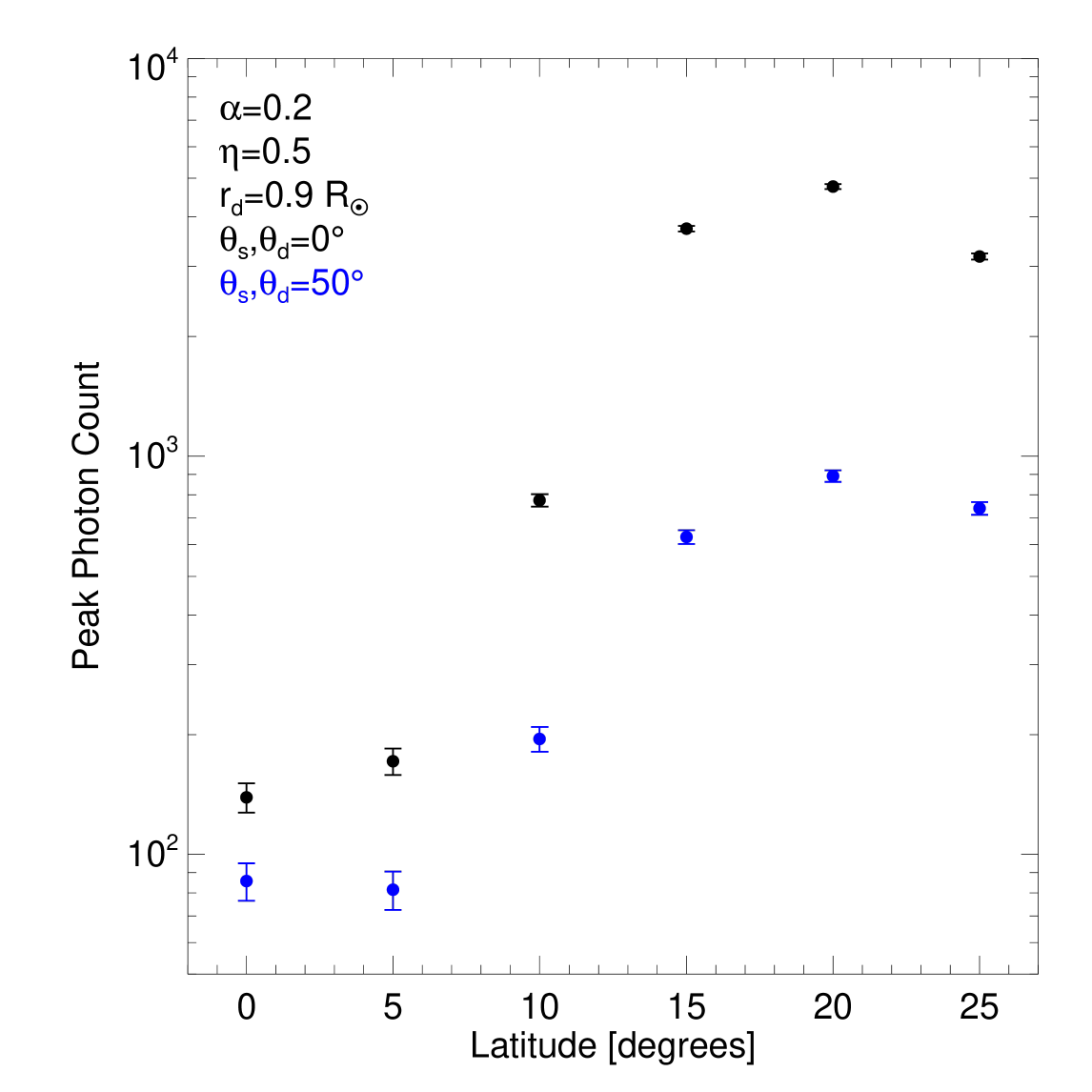}
    \caption{Peak photon counts for sources in a dipole magnetic field at polar angles between $\phi_\mathrm{s}=65\arcdeg-90\arcdeg$ (up to $25\arcdeg$ from the equator) at the disk center (black) and at $\theta_\mathrm{s}=50\arcdeg$ (blue), using $N=10^5$ photons.}
    \label{fig:peak_ph_counts}
\end{figure}

For a fixed dipole magnetic field perpendicular to the ecliptic, the local magnetic field direction varies for sources located at different polar angles $\phi$. Figure \ref{fig:decay_shift_fi_src} compares the decay times and source displacement distances across the sky-plane for two different sets of data using dipole, and radial magnetic fields. The dipole center is located at $\theta_\mathrm{d}=0\arcdeg$, $\phi_\mathrm{d}=90\arcdeg$, and $r_\mathrm{d}=0.9$~R$_\odot$, while the sources are injected at polar angles between $\phi_\mathrm{s}=65-90\arcdeg$ (up to $25\arcdeg$ above the equator) at $\theta_\mathrm{s}=0\arcdeg$. The radial case simulates the same source characteristics but with a spherically symmetric, radial magnetic field. Additionally, Figure \ref{fig:peak_ph_counts} presents the total number of collected photons for each polar source location in the dipole case. For latitudes close to $0\arcdeg$ (near the solar equator), the dipole case shows reduced source motion, prolonged decay times, and the lowest photon count. From $5-17.5\arcdeg$, the dipole case decay time reduces by $1/3$ and the source displacement decreases towards zero, contrary to the radial case which increases in distance and maintains the same decay time at all polar angles. At a latitude of $22.5\arcdeg$, the local dipole field direction matches that of the local radial direction and the characteristics are similar to the radial case. At approximately the same angle we see the largest photon count, more than an order of magnitude larger than at the equator.

\section{Solar Radio Burst Fine Structure Drift Rate}

\subsection{Convolution of a Radio Pulse with a Scattering Function}

\begin{figure*}[htb!]
    \epsscale{1.1}
    \plotone{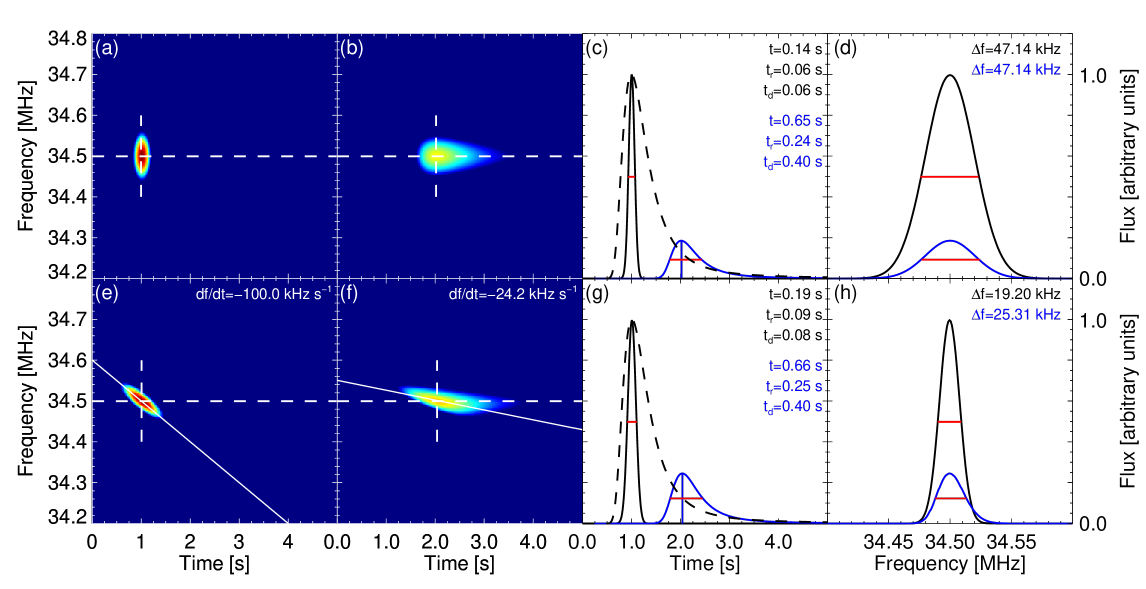}
    \caption{Simulated radio burst dynamic spectra with and without the effects of radio-wave scattering. The first column shows a radio pulse pre-scattering, assuming a Gaussian time and frequency profile. The top row (a-d) considers a burst emitted instantaneously across all frequencies, and the bottom row (d-g) consider a burst emitted with a frequency drift rate of $-100$~kHz s$^{-1}$, represented by the solid white line in panels (e). The second column shows the results of the convolution between the 2D Gaussian and the scattering function $I_s$ for $\alpha=0.25$ (dashed curve in panels (c, g)). The white dashed lines in the dynamic spectra show the locations of the time and frequency-flux profiles that cross at the maximum intensity. The profiles are displayed in the third and forth columns, respectively. The solid black curves represent the pulse Gaussian profiles, and the blue curves represent the result of the convolution. The red horizontal lines denote the FWHM.}
    \label{fig:2d_gaussian_burst_conv}
\end{figure*}

To evaluate the effect of radio-wave scattering on short duration burst fine structures observed in dynamic spectra, we consider a radio pulse in the absence of scattering effects, with intensity $I$ modeled by a 2D Gaussian of form
\begin{equation}
    I(t,f)=I_0\exp{\left[-\left(\frac{(t-t_0)^2}{2\sigma_t^2}+\frac{(f-f_0)^2}{2\sigma_f^2}\right)\right]},
\end{equation}
where $t_0$ gives the time of peak intensity $I_0$ and $f_0$ is the central frequency of the burst, with standard deviation $\sigma_{t,f}$. Figure \ref{fig:2d_gaussian_burst_conv} shows an initial pulse and the measured characteristics (decay time, frequency drift rate, and bandwidth). We set the initial duration to be $140$~ms which is consistent with the inferred duration of decameter radio spikes and striae in the absence of scattering, and a bandwidth of $47$~kHz at a central frequency of $34.5$~MHz \citep[e.g.][]{Clarkson_2021}. A Gaussian bandwidth is chosen to match the symmetrical spectral shape of fine structures \citep[e.g.][]{1993A&A...274..487C,Clarkson_2021}, and for the time profile assuming an initial Gaussian electron distribution function for plasma emission \citep{1999SoPh..184..353M,2001SoPh..202..131K,2021NatAs...5..796R}. The dashed curve in panels (c, g) represents a scattering function $I_s$---i.e. the time profile from the scattering simulations of an injected point source with a radial magnetic field from \cite{2020ApJ...905...43C} (see Figure 3c, where $\alpha=0.25$). The model used to fit the scattering function is an asymmetrical Gaussian given by
\begin{equation}\label{eq:asym_gauss}
    I_s (t)=I_0\exp{\left[-\left(\frac{t-t_0}{a+t\cdot b}\right)^2\right]},
\end{equation}
where $I_0, t_0, a, b$ are free parameters of the fit. When $b\ge 0$, the time varying standard deviation leads to a HWHM rise $t_r$ and decay $t_d$ time given by
\begin{equation}
    t_{\mathrm{r,d}} = \frac{\sqrt{\ln{2}}(a + bt_0)}{1\pm b\sqrt{\ln{2}}},
\end{equation}
for the positive and negative $b\sqrt{\ln{2}}$ terms, respectively. The second column (panels b, f) shows a convolution of the radio pulse and the scattering function $I_s$ for each frequency channel as
\begin{equation}\label{eq:convol}
    I_c(t)=\int_{-\infty}^{\infty}{I(t-\tau)I_s(\tau)}\;\mathrm{d}\tau.
\end{equation}

The convolution of the initial pulse with the scattering function mimics the time-broadening due to radio-wave scattering at each frequency, whilst leaving the total bandwidth intact. Consequently, the initial short millisecond pulse is broadened in time in dynamic spectra, and the observed frequency drift is diluted. The drift rate reduction level therefore depends on the duration of the scattering function, or physically, the turbulence level, anisotropy factor, and emitter location with respect to a non-radial field (see increased time profiles in Figure \ref{fig:decay_shift_fi_src}(a) near the loop apex). An additional effect of the convolution is that the observed bursts are delayed from the original peak time by $\sim1$~s.

\begin{figure}[htb!]
   \centering
\includegraphics[width=0.5\textwidth]{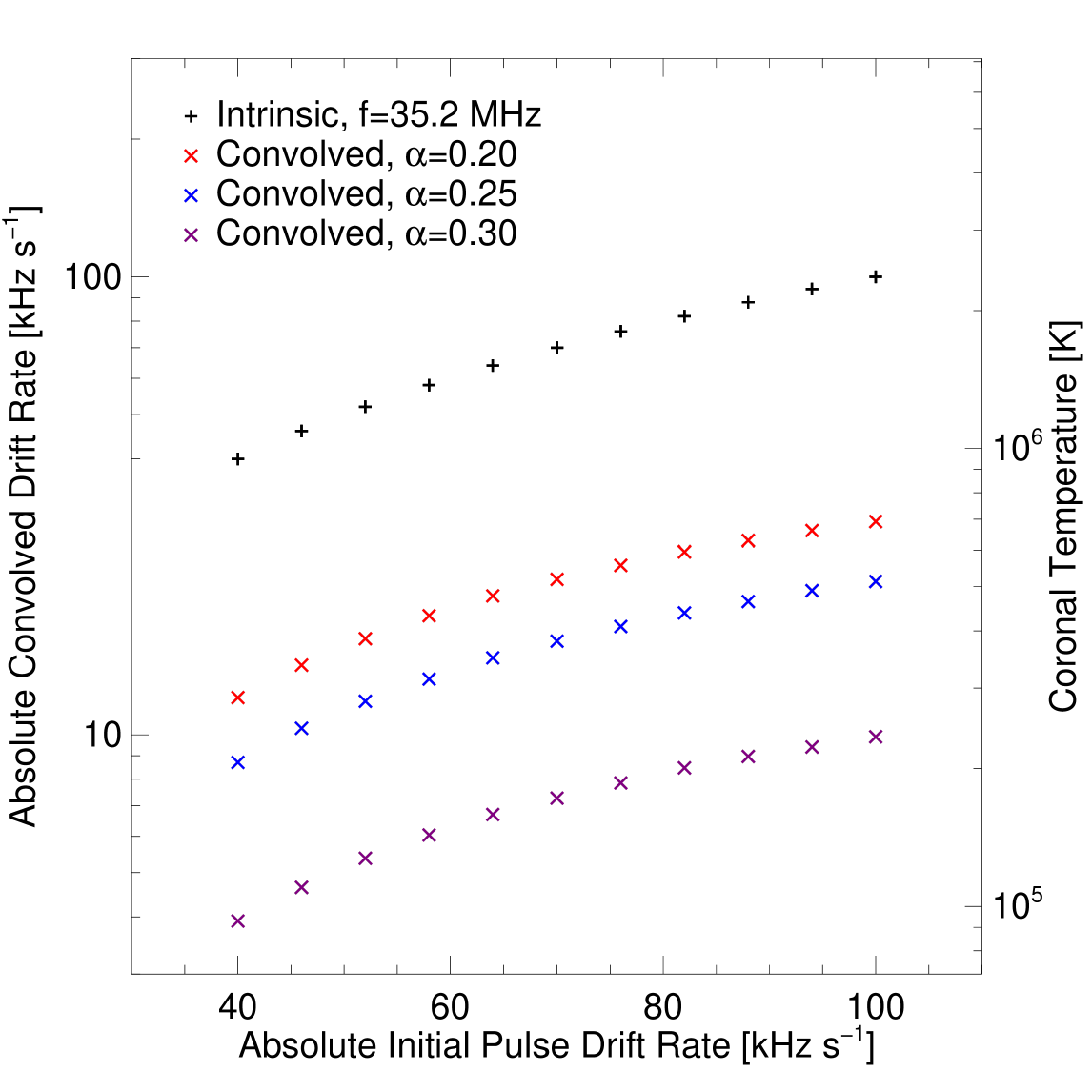}
    \caption{Frequency drift rates of the simulated initial radio pulses and after convolution with scattering function from radial magnetic field simulations using different values of $\alpha$ at 35.2 MHz.}
    \label{fig:intrinsic_2dgauss_conv}
\end{figure}

We find that an initial radio pulse with a drift rate of $-100$~kHz s$^{-1}$ is reduced to $-24$~kHz s$^{-1}$ as shown in Figure \ref{fig:2d_gaussian_burst_conv}(f). In Figure \ref{fig:intrinsic_2dgauss_conv}, the convolved drift rates are presented using scattering functions from various anisotropy values as presented in \cite{2020ApJ...905...43C} along radial magnetic fields. The initial (before convolution) fine structure drift rates are chosen such that the estimated coronal temperature is near $1-2$~MK, typical for coronal plasma. The relation between the thermal temperature and fine structure drift rate \citep{2021NatAs...5..796R} is given as
\begin{equation}\label{eq:dfdt_T}
    \frac{\mathrm{d}f}{\mathrm{d}t}=\frac{f_\mathrm{pe}}{2n}\frac{\mathrm{d}n}{\mathrm{d}r}v_\mathrm{gr}
\end{equation}
where $v_\mathrm{gr}=3v_\mathrm{Te}^2/v_b$ with $v_\mathrm{Te}^2=k_\mathrm{B}T_e/m_e$, $m_e$ and $k_\mathrm{B}$ are the electron mass and Boltzmann constant. The characteristic speed of electron beams $v_b$ is between $\sim(0.1-0.5)c$ \citep{1985srph.book..289S,2020FrASS...7...56R}, and so we take a typical speed of $v_b=0.3c$. Using the same density model as in the simulations \citep{Kontar_2019} gives $n=1.28\times10^7$~cm$^{-3}$ at $f_\mathrm{pe}=32$~MHz.
Stronger levels of anisotropy have a weaker reduction effect---for example, for an initial pulse drift rate of $-40$~kHz s$^{-1}$, the observed burst has a drift rate of $-4$~kHz s$^{-1}$ for $\alpha=0.3$, and $-12$~kHz s$^{-1}$ for $\alpha=0.2$. Using equation \ref{eq:dfdt_T}, these observed drift rates would suggest coronal temperatures of $\sim(1-3)\times10^5$~K, rather than $\sim10^6$~K. Therefore, without correcting for the drift dilution due to scattering effects, the coronal temperature would be underestimated.

\begin{figure}[htb!]
    \centering
\includegraphics[width=0.5\textwidth]{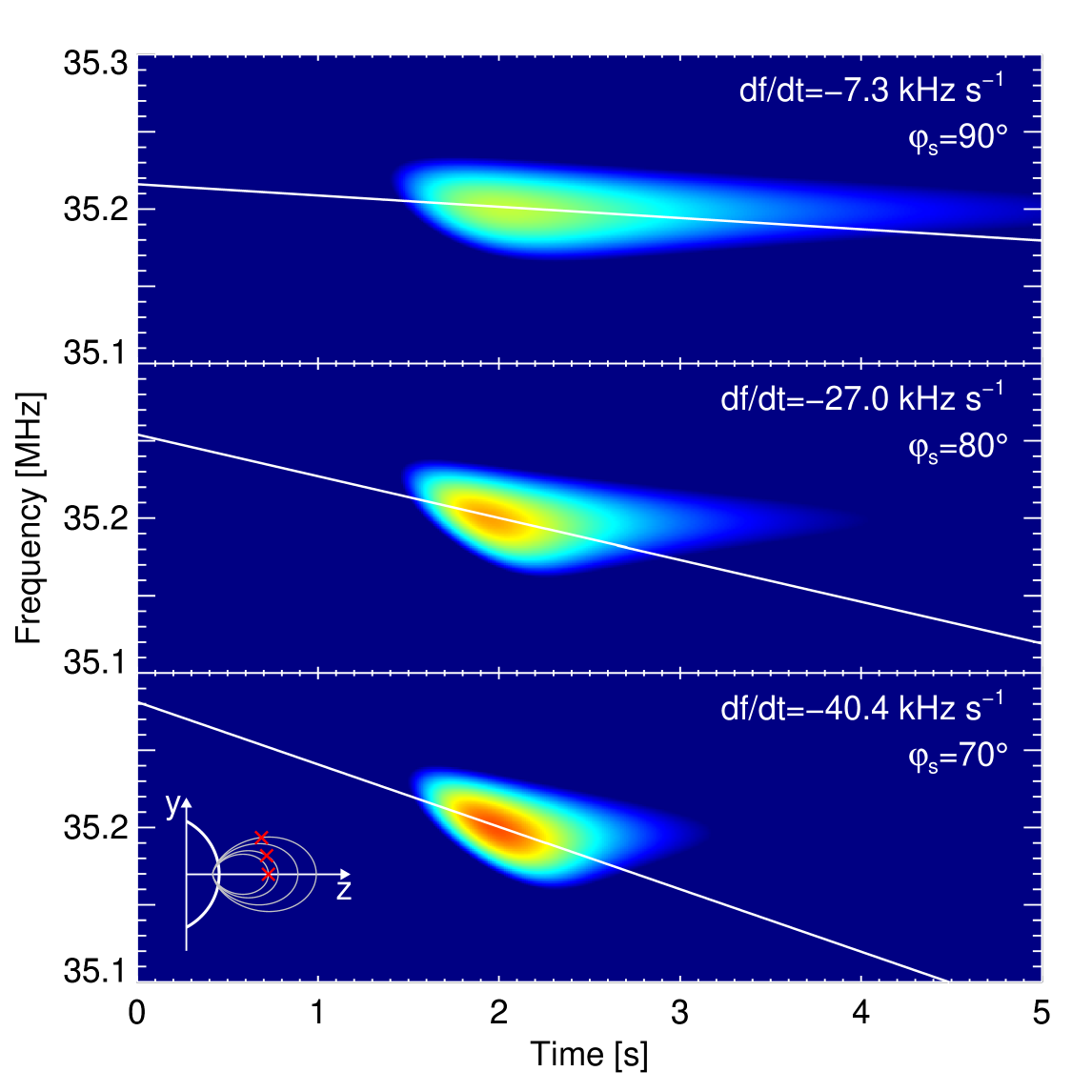}
    \caption{Dynamic spectra of drifting radio burst fine structures convolved with a scattering function for sources located at $r_\mathrm{s}=1.75$ R$_\odot$, $\theta_\mathrm{s}=0\arcdeg$, and $\phi_\mathrm{s}$ of $90\arcdeg$, $80\arcdeg$, and $70\arcdeg$ with a dipole magnetic field located at $r_\mathrm{d}=0.9$ R$_\odot$, $\phi_\mathrm{d}=90\arcdeg$ and $\theta_\mathrm{d}=0\arcdeg$, with $\alpha=0.2$ and $\eta=0.5$. The solid lines show the drift rates given by $\mathrm{d}f/\mathrm{d}t$ that decreases for $\phi_s$ farther from the equator. The inset shows the source locations (red symbols) for an observer along the +\textit{z}-axis.}
    \label{fig:2d_gaussian_burst_conv_dipolePhi}
\end{figure}

Along dipolar magnetic field lines, the scattering function is extended for sources closer to the loop apex. Figure \ref{fig:2d_gaussian_burst_conv_dipolePhi} shows three fine structures with initial pulse drift rates matching those in Figure \ref{fig:2d_gaussian_burst_conv}. Each is convolved with a scattering function simulated with a dipole magnetic field for sources located between $\phi_\mathrm{s}=70-90\arcdeg$ (up to $20\arcdeg$ from the equator). For sources near the loop apex, we find a strong drift rate reduction by 93\%, followed by weaker reduction of 73\% and 60\% for sources at $10\arcdeg$ and $20\arcdeg$ above the equator, respectively, as well as peak intensity reduction closer to the loop apex.

\section{Summary}

The introduction of a non-radial magnetic field to anisotropic radio-wave propagation simulations unveils the significance of scattering effects combined with the projection of the magnetic topology in shaping the apparent source motion and fine structure morphology.

The non-radial nature of the dipole magnetic field 
means that the sources are broadened 
along an axis different to the spherically symmetric radial magnetic field.  
For a source at the disk center with the magnetic field 
oriented towards the solar north, 
the resulting sky-plane apparent source expansion is along the \textit{x}-axis. Over time, the photons escape the scattering region preferentially parallel to the magnetic field, 
manifesting as a shift vertically in the sky-plane. On the other hand, a radially symmetric magnetic field would produce stationary expansion with a maintained circular shape \citep[see][Figure 1]{2020ApJ...905...43C}. Additionally, the time profile and image highlight the emergence of the radio echo, 
comparable to those observed in drift-pair bursts, 
although not as intense \citep{2020ApJ...898...94K}. 
The lower flux of the echo in a non-radial field could be caused by the photons scattered towards the plasma frequency surface at an indirect angle due to the field direction, thereby reducing the number of photons that are reflected towards the observer. 
Despite this, the echo noticeably changes the image morphology by increasing the apparent size and inciting a shift of the imaged source due to the combined location of two spatially separated components (Figure \ref{fig:sim_images_TH_FI_0}), 
therefore producing an abrupt change of the source trajectory 
and position when imaging along the burst tail.

For initial sources located at the loop apex, the field direction is parallel to the plasma frequency surface. The photons have no preferential escape direction from the scattering region, propagating in either direction along the field. Consequently, this results in a fixed centroid location over time representing the average position of an extended apparent source (Figure \ref{fig:fullSun_radial_dipole_vectors}(b)). Moreover, the photons remain near the plasma frequency for longer and experience increased scattering such that the decay times are prolonged, 
and the probability of absorption is increased. Therefore, the model predicts that radio emission emanating from magnetic regions parallel to the plasma frequency surface 
may appear fainter with increased duration than sources along radial, open field lines. By not accounting for the extended absorption period caused by scattering in complex magnetic regions, one might mistakenly assume the presence of weaker intrinsic sources.

Importantly, the vertical sky-plane projected source motion of solar radio spikes and Type IIIb striae observed by LOFAR \citep{Clarkson_2021, 2023ApJ...946...33C} is found
in agreement with their interpretation of strong anisotropy 
in post-flare, extended coronal loops of $\alpha=0.2$. 
The $(0.1-0.2)$~R$_\odot$ sky-plane separation between the injection location and imaged source height implies that the emitter location is not observed in radio images, consistent with the radial simulations presented in \cite{2023ApJ...956..112K}. Further, in Figure \ref{fig:rdi09_Fisrc-5_gammaRot}, we reproduce the observed source motion of a Type IIIb stria towards the solar disk as shown in \cite{Clarkson_2021} by rotating the magnetic field structure to align between the active regions observed by AIA. The figure also shows a potential-field source-surface extrapolation for the event. Whilst Type IIIb sources are excited along open field lines, the temporal trajectory of the centroids at a given frequency implies radio-wave propagation that is dominated by scattering in a non-radial magnetic field. The result shows that peculiar, fixed frequency source trajectories observed in radio emission can be caused by the projection of a complex magnetic field structure onto the sky-plane in an anisotropic scattering medium. Additionally, due to a non-radial local magnetic field, we find that the simulated sources have a different shape, size, and orientations. Interestingly, the apparent source sizes could be smaller than that in the radial magnetic field geometry. Consequently, in order to correctly interpret the apparent source motion in radio images, the time along the burst, source location, magnetic field structure, and plasma turbulence should be considered simultaneously.

\subsection{Apparent source Bifurcation}

An interesting result from the simulations is that environments characterized by a strong anisotropy factor of $\alpha=0.1$ with a magnetic field that is near parallel to the plasma frequency surface (a loop-like structure), 
the apparent source can appear to bifurcate into two separate components (Figure \ref{fig:split_source}) that each shift in opposite directions. The appearance of two separated sources is linked to the photons escaping the scattering region preferentially in either direction along the guiding field, and is particularly noticeable in the strongest anisotropy case that offers high directivity. In Figure \ref{fig:split_source}, the photons are emitted at $-5\arcdeg$ from the dipole apex where the plasma frequency surface is near parallel to the field direction. This emission site allows the photons to scatter along the magnetic field in both directions; contrary to a radial field in which the photons propagating towards the Sun are reflected outward. The photons that are preferentially scattered down the dipole leg encounter the plasma frequency surface after propagating a small distance and are subsequently reflected, causing this secondary component to halt in position. These photons remain in the scattering region for longer allowing for increased absorption such that the lower of the two components appears fainter. Figure \ref{fig:split_source} shows a separation distance between the two components at 35 MHz of a few hundred arcsec at a time they share a similar brightness, requiring an angular resolution $\leq3$~arcmin to spatially resolve. Similar bidirectional, vertical source motion across the sky-plane has been observed previously above 80 MHz \citep{2017ApJ...851..151M}, yet is explained as the result of multiple injection of electron beams along diverging \textit{open} field lines, with the source splitting motion at individual frequencies interpreted as a projected time-of-flight effect. Our model suggests that a single emitter in \textit{closed} field lines can feasibly appear as multiple apparent sources due to strong anisotropy and non-radial magnetic fields along the line of sight.

\subsection{Influence of Scattering on Fine Structure Drift Rates}

The scatter broadening of a radio pulse presents a number of notable effects. The resulting time profile is broadened, and accompanied by a delay of the peak time, as well as a reduction in the fine structure drift rate. The temporal broadening of the pulse at each frequency results in a smearing of the emission in dynamic spectra, leading to the delay of the peak by $\sim 1$~s at decameter frequencies, as shown by \cite{2023MNRAS.tmp..341C} for fundamental emission. Whilst we convolve the same scattering function at each frequency channel that is occupied by the initial pulse, the narrow bandwidths encompass a frequency range across which the scattering rate would not change significantly.

Variations in the degree of anisotropy present in the intervening plasma influence the observed drift rate, with a more pronounced reduction evident under conditions closer to isotropy. Moreover, the observed drift rate of sources emitted in an anisotropic plasma can be subject to further alteration due to the magnetic topology, dependent on the source location within it due to a varied scattering contribution (Figure \ref{fig:2d_gaussian_burst_conv_dipolePhi}). The recent association of the fine structure drift rate with the Langmuir wave group velocity and coronal temperature \citep{2021NatAs...5..796R} means that the observed drift rate underestimates these quantities. For instance, without a scattering correction, the median fine structure drift rates of spikes associated with a coronal loop near 30 MHz observed by \cite{2023ApJ...946...33C} were $|\mathrm{d}f/\mathrm{d}t| \sim (10-20)$ kHz s$^{-1}$, inferring $T_e\sim (0.3-0.6)$~MK, rather than $(1-1.5)$~MK. In a turbulent medium with anisotropy of $\alpha=0.2$, such observed drift rates imply $(40-70)$~kHz~s$^{-1}$ when decoupled from the scattering effects. However, we stress further that the scattering contribution in complex magnetic fields is also dependent on the source location within the structure, which should be considered alongside the turbulent properties of the plasma in order to interpret the fine structure morphology.

If we consider fine structure drift rates at higher frequencies \citep[e.g.][]{1990A&A...231..202G, 2005A&A...434.1139D, 2016A&A...586A..29B} where the coronal density gradient is steeper, one would anticipate higher fine structure drift rates for a source propagating at a given speed. Indeed, observations between 200 MHz to 4 GHz of radio burst fine structures do present larger drift rates than those observed in the decameter range (Figure \ref{fig:fineStructure_dfdt}). However, the trend exhibited by these high frequency structures is closer to what Type III bursts present below 550 MHz \citep{1973SoPh...29..197A}, and not necessarily consistent with the drift rates of decameter fine structures. The reason is not entirely clear---firstly, the fine structure bandwidths above 400 MHz are approaching the instrumental spectral resolutions \citep[see Figure 11 in][for example]{2023ApJ...946...33C}, which means that the measured drift rates may over-estimate the average at these coronal heights. Secondly, the turbulence model used decreases towards the Sun from a peak near $(4-7)$~R$_\odot$. This decrease leads to shorter time profiles accompanied by a weaker suppression of the drift rate. Consequently, it's not clear whether these high frequency drift rates should continue the trend set by the decameter fine structures, or depart from it due to a difference in the scattering regime. Nevertheless, the effect of scattering should be accounted for in both frequency domains, although it is likely weaker for decimeter bursts.

\begin{acknowledgments}
The work was supported by the Dstl funding through the UK-France PhD Scheme (contract DSTLX-1000106007) and by STFC/UKRI grants ST/T000422/1 and ST/Y001834/1.
We thank Nicolina Chrysaphi, Mykola Gordovskyy, and Francesco Azzollini for helpful discussions.
\end{acknowledgments}

\bibliographystyle{aasjournal}
\bibliography{main}

\appendix
\restartappendixnumbering

\section{Decameter \& Decimeter Fine Structure Drift Rates}

\begin{figure}[htb!]
    \epsscale{0.7}
    \plotone{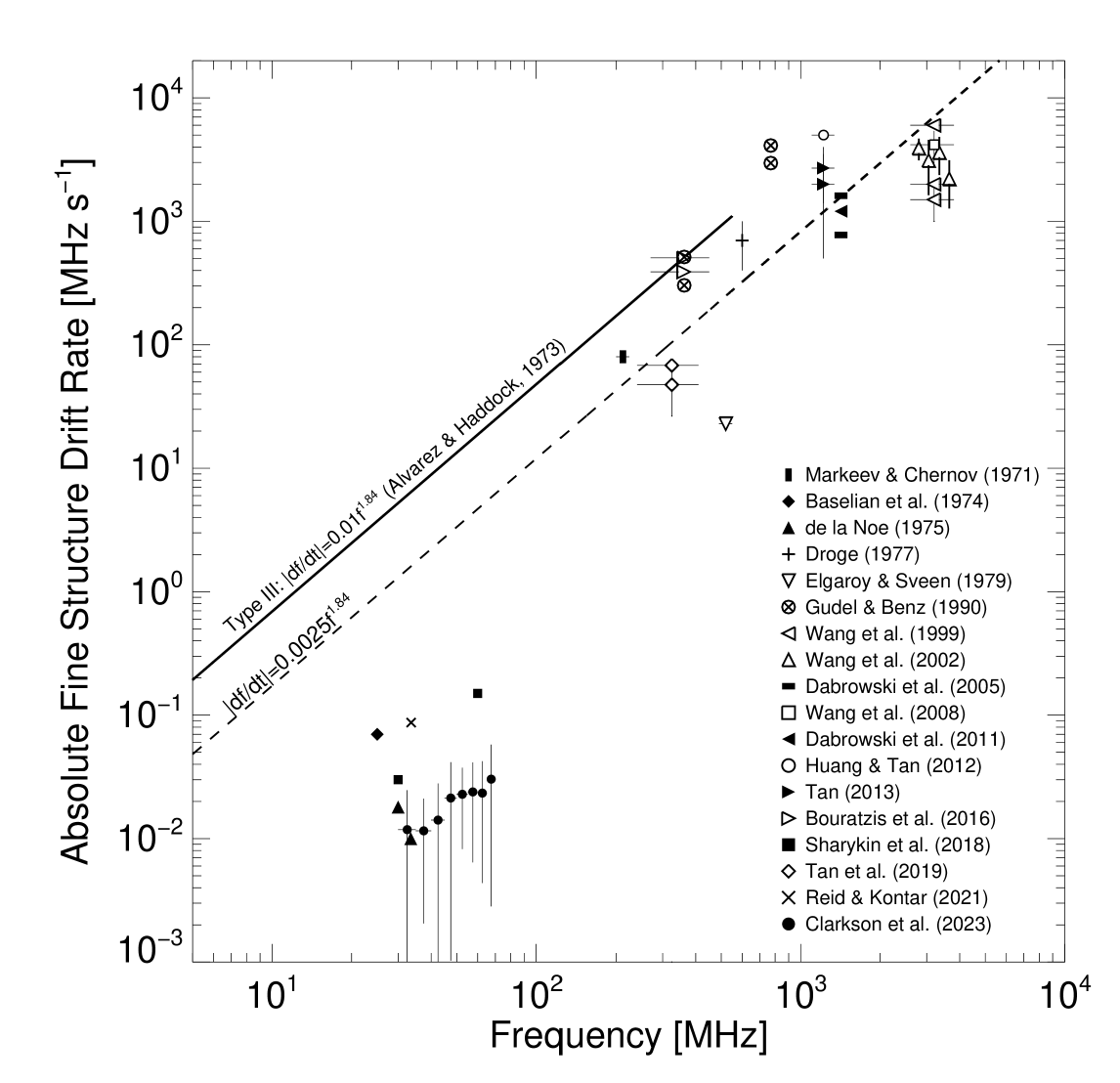}
    \caption{Compilation of radio burst fine structure average drift rates from numerous studies \citep{1971SvA....14..835M, 1974SoPh...39..213B, 1975A&A....43..201D, 1977A&A....57..285D, 1979Natur.278..626E, 1990A&A...231..202G, 1999SoPh..189..331W, 2002SoPh..209..185W, 2005A&A...434.1139D, 2008SoPh..253..133W, 2011SoPh..273..377D, 2012ApJ...745..186H, 2013ApJ...773..165T, 2016A&A...586A..29B, 2018SoPh..293..115S, 2019ApJ...885...90T, 2021NatAs...5..796R, 2023ApJ...946...33C}. The solid line shows the Type III drift rate relation by \citep{1973SoPh...29..197A}. The dashed line shows the same slope offset to match the data above 200 MHz. The uncertainty for \cite{2023ApJ...946...33C} represents the interquartile range for the median data.}
    \label{fig:fineStructure_dfdt}
\end{figure}

Figure \ref{fig:fineStructure_dfdt} shows the fine structure drift rates from different studies, observed across three decades in frequency. For fine structures above 200 MHz, \cite{2023ApJ...946...33C} (see Figure 12) show that the bandwidth measurements of spikes at these frequencies are approaching the limits of the instrumental spectral resolution. Correspondingly, it would be challenging to measure an accurate frequency drift rate for such structures, and we expect that these values could be overestimating the average.

\end{document}